\documentclass[twocolumn]{aastex701}

\usepackage{amsmath}
\usepackage{commath}
\usepackage{comment}
\usepackage{cleveref}
\usepackage{subcaption}
\usepackage{overpic}

\begin{document}

\title{Microwave Polar Brightening and Its Connection to Polar Coronal Holes}

\correspondingauthor{Anshu Kumari}
\email{anshu@prl.res.in}


\author[orcid=0009-0009-1355-5631,gname=Rohan, sname=Bose]{Rohan Bose}
\affiliation{Aryabhatta Research Institute of Observational Sciences, Nainital-263002, Uttarakhand, India}
\affiliation{Department of Physics, Indian Institute of Technology Roorkee, Roorkee-247667, Uttarakhand, India}
\email{rohannewone@gmail.com}  

\author[orcid=0000-0001-5742-9033,gname=Anshu, sname=Kumari]{Anshu Kumari} 
\affiliation{Udaipur Solar Observatory, Physical Research Laboratory, Dewali, Badi Road, Udaipur-313 001, Rajasthan, India}
\email{anshu@prl.res.in}

\author[orcid=0000-0002-6954-2276,gname=Vaibhav,sname=Pant]{Vaibhav Pant}
\affiliation{Department of Physics, Indian Institute of Technology, Delhi, New Delhi 110016, India}
\email{vaibhavpant55@gmail.com}

\author[orcid=0009-0008-5834-4590,gname=Srinjana,sname=Routh]{Srinjana Routh}
\affiliation{Aryabhatta Research Institute of Observational Sciences, Nainital-263002, Uttarakhand, India}
\affiliation{Department of Applied Physics, Mahatma Jyotiba Phule Rohilkhand University, Bareilly-243006, Uttar Pradesh, India}
\email{srinjana.routh@gmail.com}

\author[orcid=0009-0001-7689-0084, gname=Divya,sname=Paliwal]{Divya Paliwal}
\affiliation{Udaipur Solar Observatory, Physical Research Laboratory, Dewali, Badi Road, Udaipur-313 001, Rajasthan, India}
\affiliation{Indian Institute of Technology Gandhinagar, Palaj, Gandhinagar, Gujarat, India}
\email{divya@prl.res.in}

\author[orcid=0000-0002-5359-5592,gname=Sunil Krishna, sname=M.V.]{M. V. Sunil Krishna}
\affiliation{Department of Physics, Indian Institute of Technology Roorkee, Roorkee-247667, Uttarakhand, India}
\affiliation{Centre for Space Science and Technology, Indian Institute of Technology Roorkee, Roorkee, India}
\email{mv.sunilkrishna@ph.iitr.ac.in}  


\begin{abstract}
Polar brightening (PB) observed at microwave frequencies serves as an important probe to study the thermal and magnetic properties in the Sun’s polar regions. Building on earlier studies that linked microwave PB to polar faculae, small-scale loops, and the polar coronal holes (PCHs), we present a comprehensive analysis of the long-term behaviour of 17 GHz microwave PB and its relation to polar magnetic field and coronal hole evolution. Using daily Nobeyama Radioheliograph observations spanning 1992 to 2018, we quantify microwave PB peak temperature variations and compare them with the temporal evolution of PCH area extracted from SDO/AIA-based SPoCA coronal hole catalogues during the period 2010-2018. We also examine the correspondence between microwave PB and the polar magnetic field to assess the nature of their association. Our results show a strong correlation between microwave PB peak temperature and PCH area, as well as with the polar magnetic-field strength. In addition, we found that regions of enhanced microwave emission are frequently associated with small-scale loop structures, consistent with Coronal Bright Points (CBPs), which are often associated with the eruption of jets. Overall, this study aims to investigate the impact of coronal holes, polar magnetic fields, and small-scale polar activity on polar brightening observed at 17 GHz and its long-term evolution. 
\end{abstract}


\section{Introduction}\label{sec:introduction}
The Sun looks very different depending on the wavelength at which it is observed, because the dominant emission processes change across the electromagnetic spectrum. In the ultraviolet range (10-400 nm), most of the observed radiation originates from strong spectral line emission formed in the chromosphere and corona by different ionised atoms. In contrast, at radio wavelengths (1 mm-100 Mm), strong spectral lines are generally absent in the case of the Sun due to the high densities in the lower solar atmosphere leading to severe pressure broadening, which effectively suppresses narrow line features \citep{Nindos2020}. As a result, radio and microwave emission from the Sun is primarily produced by three incoherent mechanisms: thermal bremsstrahlung (free–free) emission, which dominates the quiet Sun and non-flaring active regions; gyroresonance emission originating from hot coronal plasma at temperatures of order $10^{6}$ K; and gyrosynchrotron radiation generated by mildly relativistic electrons in flare environments \citep{Nindos2020}.

Within this microwave regime, one of the most prominent features on the Sun is polar brightening (PB, hereafter), also referred to as polar limb brightening, polar-cap brightening, or enhanced temperature regions in the literature. These bright patches near the solar poles are characterised by a brightness temperature ($T_B$) that exceeds the surrounding values by approximately $3–20 \% $ (depending on the observing wavelength; see \citealt{Selhorst2003}), and are thought to be associated with features intrinsic to poles on top of limb brightening due to the temperature gradient in solar atmosphere \citep{Shibasaki1998}. These brightenings were first reported by \citet{Babin1976} in millimetric and centimetric raster-scan observations. The photospheric features, which are thought to be associated with the microwave PBs, are the white light polar faculae (\citealt{Riehokainen2001, Gelfreikh2002, Selhorst2003}). Since polar faculae may host magnetic fields reaching kilogauss strengths \citep{Homann1997}, \citet{Riehokainen2001} proposed that the brightenings might contain a gyrosynchrotron component. Similarly, when discussing the chromosphere, the microwave PBs are suggested to be related to the bare/ hole regions in the spicular forest \citep{Selhorst2005a, Selhorst2005b} lying above polar faculae. Most of these interpretations are supported by the observed correlation between polar brightening and the underlying polar magnetic field \textbf{\citep{Gopalswamy2012,Shibasaki2013}}. \citet{Nitta2014} suggested that the solar cycle variation of $T_{b}$ may broadly reflect that of the coronal holes from the correlation between  NoRH 17 GHz  microwave PB and the polar magnetic field, and comparing NoRH and AIA 171  \AA{} daily maps (also see \citealt{Selhorst2010}, which explored the correlation between NoRH 17 GHz and EIT/ SOHO EUV temporal intensity profiles at the poles). There have been multiple other studies trying to find if the Polar Coronal Holes (PCHs) can be the coronal signatures of the microwave PBs. \citet{Silva2016} utilised microwave and EUV limb synoptic charts (also known as Butterfly diagrams) to demonstrate the coexistence of microwave PBs and PCHs (also see \citet{Selhorst2003} and \citet{Kim2017}). However, the correlation coefficient they found was low (0.36 and 0.39, respectively, for the North and South poles between 17 GHz and 171 \AA{} maps). Extending these studies, \citet{Fujiki2019} compared the temporal evolution of NoRH 17~GHz microwave PB in the period 1992–2017 with PFSS-derived coronal hole area (from polar magnetic field distribution) and found a strong correlation (CC~$\approx$~0.97 for both the poles) between microwave PB and PCH regions. However, it should be noted that PFSS-derived coronal hole areas describe the large-scale magnetic topology, whereas coronal holes identified in EUV observations are governed by the thermodynamic structure of the corona that determines the EUV emissivity \citep{Selhorst2003, Silva2016, Kim2017}. 

In the present study, we aim to analyse the long-term variation of the 17 GHz microwave PB across different phases of the solar cycle and examine its relationship with the temporal evolution of polar coronal hole areas derived directly from EUV observations. We also seek to investigate how the polar magnetic field varies with the solar cycle and how this variation differs in relation to the area of the polar coronal hole. In addition, we will perform one-to-one comparisons between individual 17~GHz microwave polar brightening patches and their corresponding EUV features to identify which EUV structures, if any, correspond to the microwave brightenings. 



\section{Observation} \label{sec:observation}

The Nobeyama Radioheliograph (NoRH), which operated from July 1992 to March 2020, observed the whole solar disc every 1 s at frequencies of 17 and 34 GHz (\citet{Nakajima1994}). It consists of eighty-four 80-cm-diameter parabolic antennas arranged in a Tee-shaped array extending 490 m in east-west and 220 m in north-south directions, with antennas situated at d,2d,4d,8d, and 16d distances from the centre of the T, where d is the fundamental baseline of length about 1.5 m. The best spatial resolution is approximately 10\arcsec\ at 17 GHz and 5\arcsec\ at 34 GHz, depending on the ratio of the observation wavelength to the maximum antenna spacing. In the current study, we have used the cleaned and aligned solar full-disc data from the 17 GHz channel, with dimensions of $512 \times 512$ pixel$^2$ and a pixel size of 4.91\arcsec. In most analyses, we have used all the available daily data from the NoRH 17 GHz band (July 1, 1992 - March 31, 2020). Still, for some parts, we used the data up to December 31 2017, as the quality of the later phases of the data had substantially deteriorated.  

For obtaining the long-term variation in coronal hole area, we used the coronal hole boundary coordinates provided by the Heliophysics Events Knowledgebase (HEK). These coordinates are derived from near–real-time observations taken by the Atmospheric Imaging Assembly (AIA) onboard the Solar Dynamics Observatory \citep{Lemen2012}, using the Spatial Possibilistic Clustering Algorithm (SPoCA; \citealt{Verbeeck2014}). The SPoCA algorithm detects coronal holes in AIA 193~\AA{} images, a wavelength particularly sensitive to coronal plasma at temperatures of approximately 1.5~MK, where coronal holes exhibit the highest contrast relative to the surrounding corona. For this analysis, we employed daily coronal hole maps spanning the period from 19~May~2010 to 18~June~2025. Data beyond March 2020 were used only to examine the long-term behaviour of coronal hole area, as no Nobeyama data are available beyond March 2020.

To investigate magnetic field variability, we utilize long-term radial magnetic field synoptic maps from the Wilcox Solar Observatory (available as tabulated data on its official website \footnote{\url{[http://wso.stanford.edu]}}) covering 1992–2020. These synoptic maps have an effective cadence of $\sim$9 hr. We also employ polar magnetic field measurements from the same observatory to examine the relationship between polar field strength and microwave PB intensity; these data have a cadence of $\sim$10 days.


\section{Analysis} \label{sec:analysis}

\subsection{Temporal variation of the polar brightening peak temperature} \label{subsec:t_v_p_b}
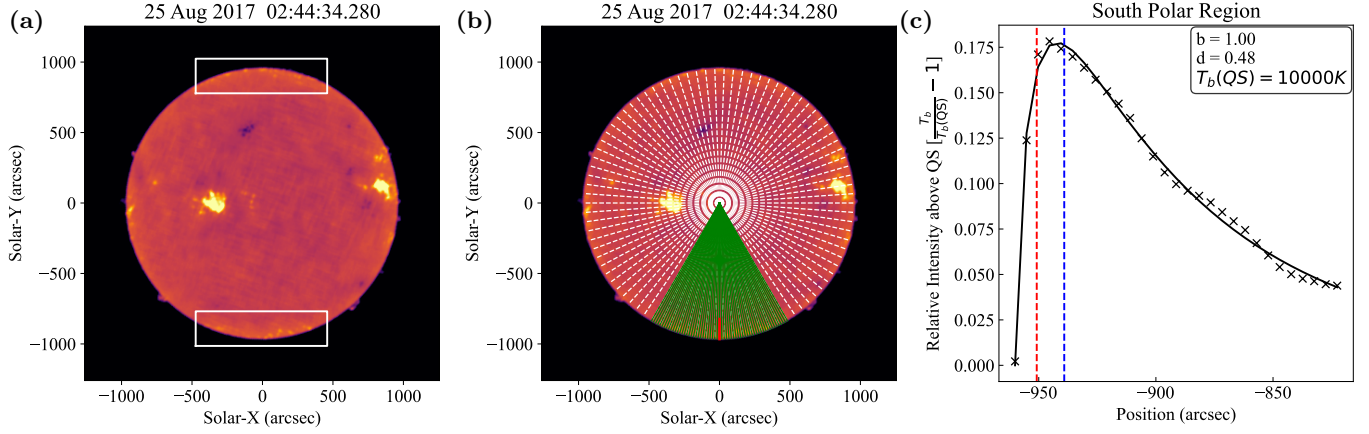
\begin{figure*}[ht!]
    \centering
    \setlength{\unitlength}{1\textwidth}
    \begin{picture}(1,0.33)
        \put(0,0){\includegraphics[width=1\linewidth]{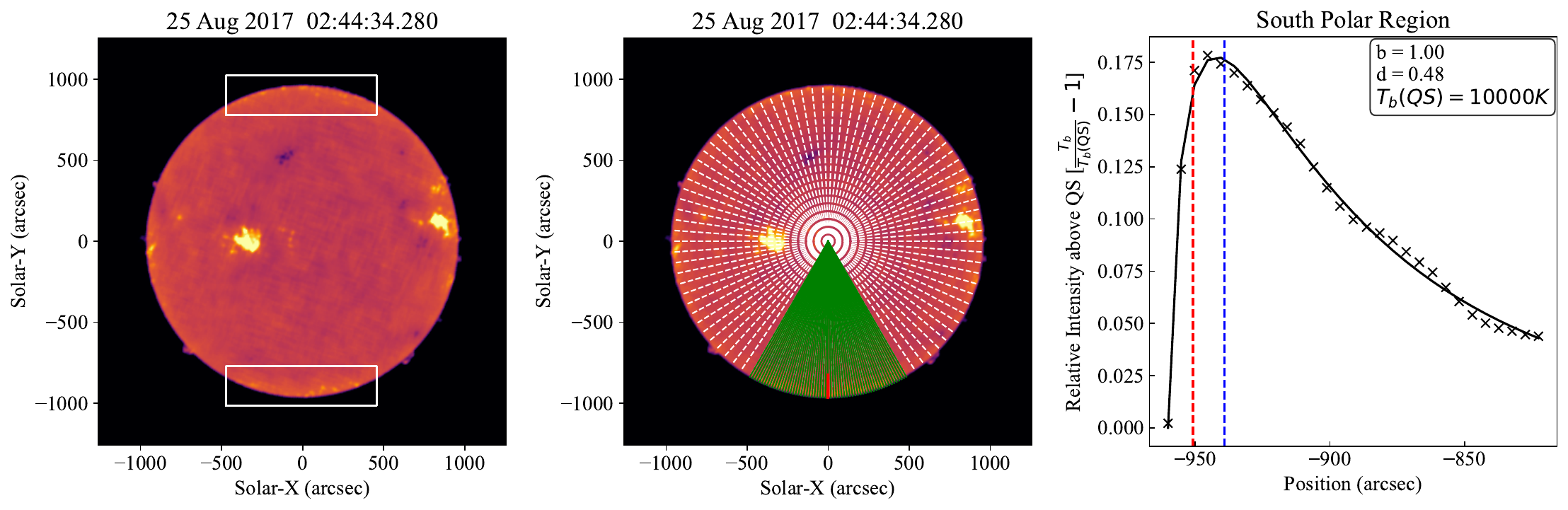}}
        \put(0.007,0.30){\textbf{(a)}}
        \put(0.335,0.30){\textbf{(b)}}
        \put(0.66,0.30){\textbf{(c)}}
    \end{picture}
    \caption{(a) \textbf{17 GHz} microwave map taken by Nobeyama Radioheliograph on 10th October 2017. White boxes highlight the regions of the microwave PBs. (b) Sun centre to limb lines drawn over the microwave map with white dashed lines at $4^{\circ}$ intervals. (c) Average 60 profiles $\pm 30^{\circ}$ about the solar south pole. Green solid lines in (b) show the positions from where radial profiles are taken, and the red small line shows the extent of the x-axis in (c). The red vertical dashed line shows the position of the solar limb. The blue one shows the limb's position in visible light (all limb and contour positions in \autoref{fig:5} are plotted accordingly).}
    \label{fig:1}
\end{figure*}

\begin{figure}[h!]
    \centering
    \setlength{\unitlength}{1\textwidth}
    \begin{picture}(1,0.37)
        \put(0,0){\includegraphics[width=1\linewidth]{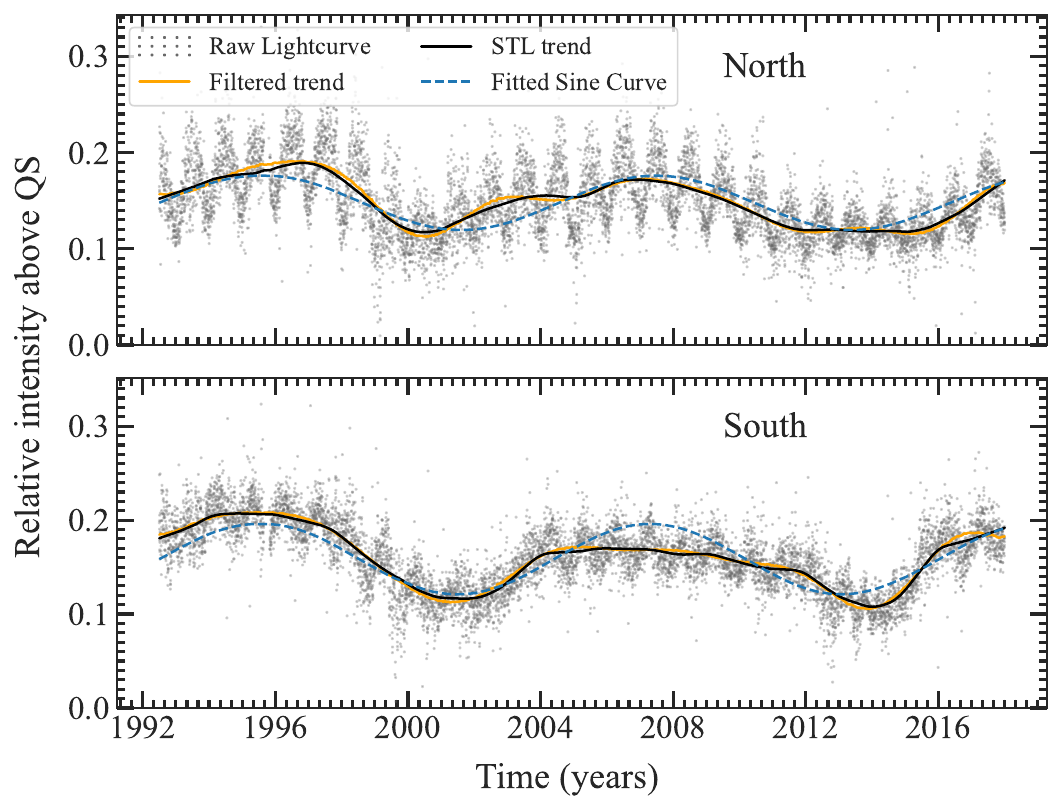}}
        \put(0.43,0.33){\textbf{(a)}}
        \put(0.43,0.17){\textbf{(b)}}
    \end{picture}
    \caption{Temporal variation of the polar brightness in the north (a) and south (b) poles. Grey dots denote daily values, orange curves represent smoothed Fourier-filtered data, black curves indicate the long-term background trends, and blue curves show the fitted sinusoidal variations.}
    \label{fig:2}
\end{figure}

In \autoref{fig:1}(a), we show a 17 GHz microwave image of the Sun obtained with the Nobeyama Radioheliograph, with the microwave PB regions marked by white boxes. The method of analysis used in this study is similar to that described in detail in \cite{Selhorst2003}; however, a brief description is provided below. For each daily map, we extracted radial intensity (brightness temperature, $T_B$) profiles at $1^{\circ}$ intervals, beginning at the west limb and proceeding anticlockwise around the disk (\autoref{fig:1}(b)). For each pole, the profiles within $\pm 30^{\circ}$ of the pole were averaged to obtain a representative radial profile for that day. As our focus is exclusively on the microwave PBs, only the portion of each profile corresponding to latitudes above $\abs{60^{\circ}}$ was retained. The choice of $\abs{60^{\circ}}$ ensures that contamination from active regions, typically confined within $\approx \abs{50^{\circ}}$ latitude is minimised, while also accommodating the annual variation of the $B_{0}$ angle (within $\pm 7^{\circ}$). Each daily radial profile was fitted using the functional form:
\begin{equation}
f(x) = a x^{b} e^{-c x^{d}},
\end{equation}
where $f(x)$ denotes the brightness temperature at radial position $x$ (in pixels), with $x$ increasing from the limb toward the disc centre (\autoref{fig:1}(c)), and a,b,c,d are all fitting parameters. The peak value of the fitted function was considered as the microwave PB peak temperature. The resulting daily peak values for the north and south poles are shown as grey points in \autoref{fig:2} (data points exceeding $4\sigma$ from the mean were excluded). To remove periodic fluctuations associated with the annual and semi-annual variation of the $B_{0}$ angle, we applied a Fourier filter retaining only periods longer than one year, followed by a running average to obtain the smoothed trend (orange curves in \autoref{fig:2}). In addition, the long-term background variation was derived using the Seasonal-Trend Decomposition by LOESS (STL) technique (shown by the black curves in \autoref{fig:2}, see~\autoref{sec:appendixA} and \citealt{Robert1990}). Finally, a sinusoidal model was fitted to the unfiltered daily peak values to characterise the cyclic behaviour, and is overplotted in blue in \autoref{fig:2}.

\subsection{Temporal variation of polar coronal hole area}\label{subsec:t_v_p_ch} 
\begin{figure*}[!ht]
    \vspace*{-0.17\textwidth}
    \centering
    \setlength{\unitlength}{1\textwidth}
    
    \makebox[\textwidth][c]{%
        \begin{picture}(0.46,0.46)
            \put(0,0){\includegraphics[width=0.456\linewidth]{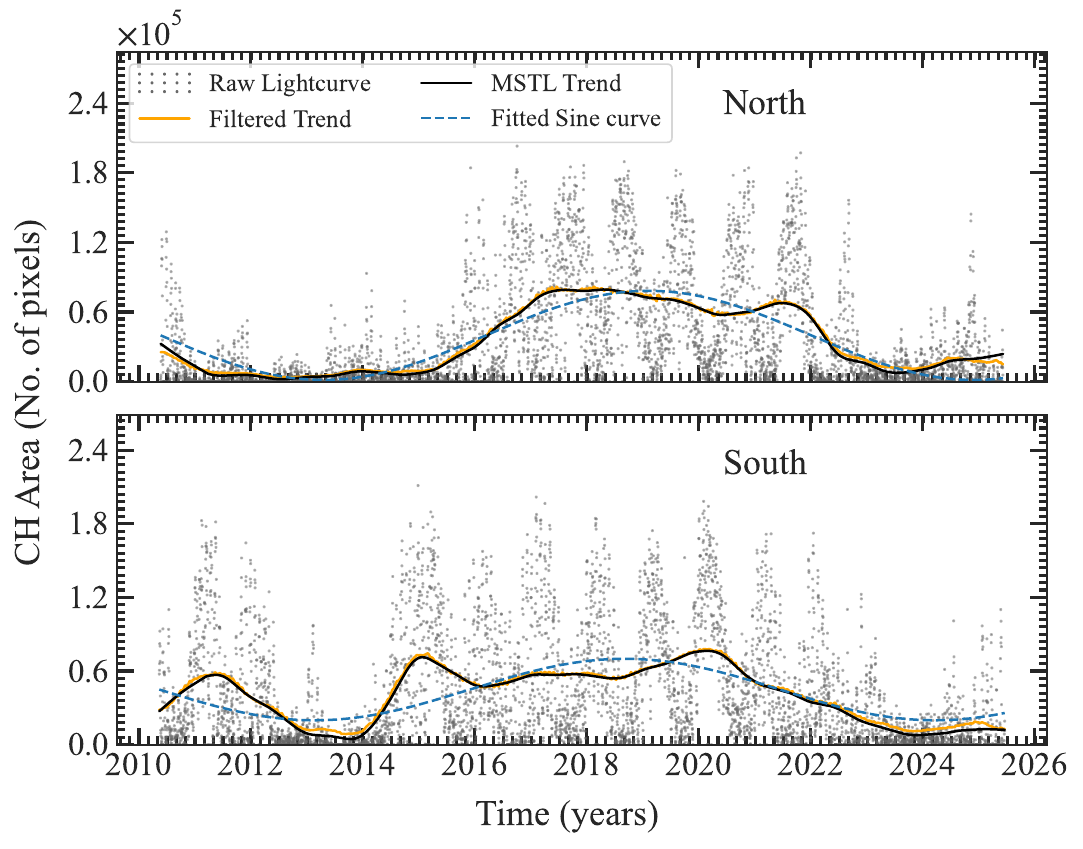}}
            \put(0.4,0.31){\textbf{(a)}}
            \put(0.4,0.16){\textbf{(b)}}
        \end{picture}
        \hspace{0.015\textwidth}
        \begin{picture}(0.50,0.54)
            \put(0,0){\includegraphics[width=0.5\linewidth]{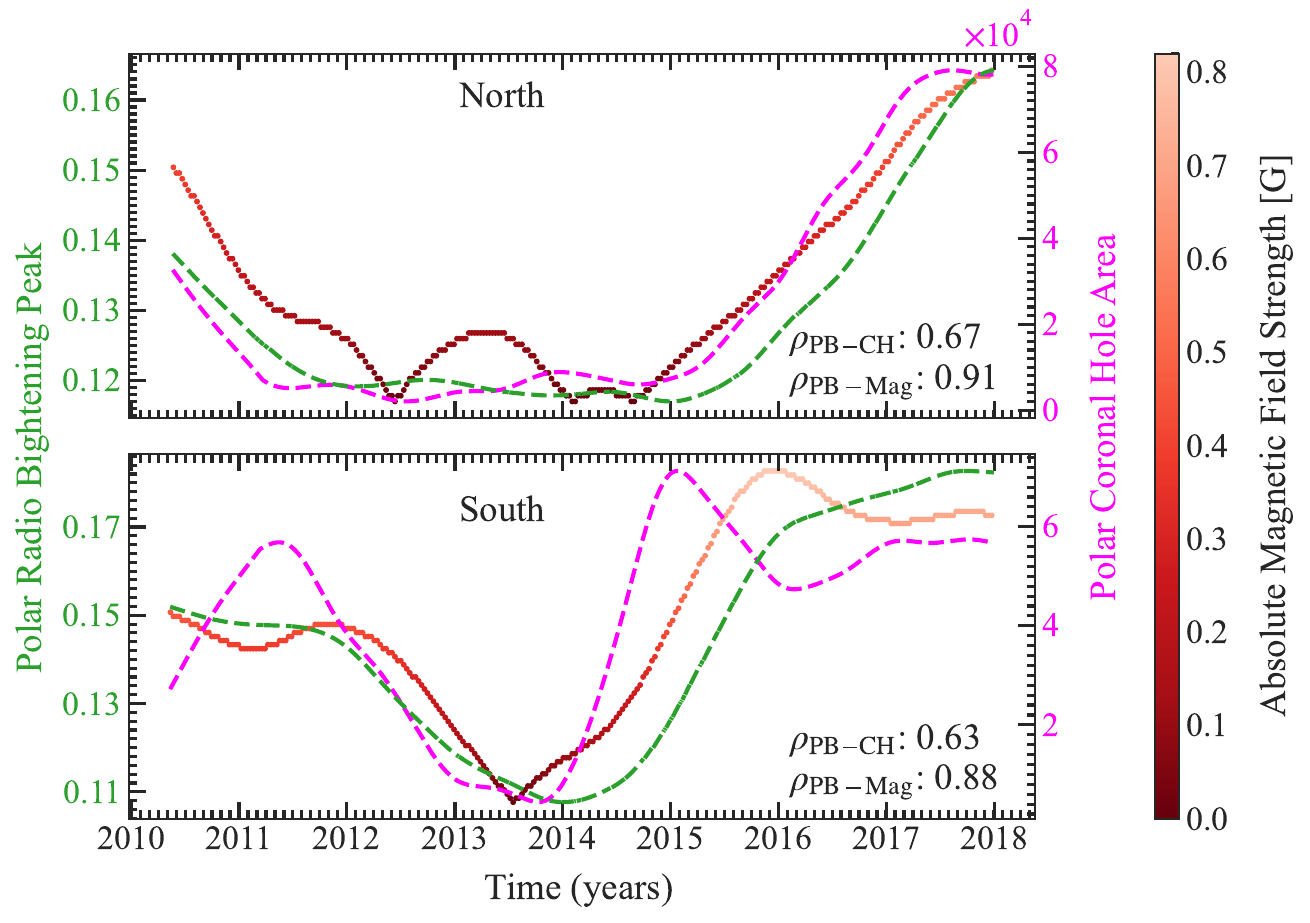}}
            \put(0.37,0.3){\textbf{(c)}}
            \put(0.37,0.15){\textbf{(d)}}
        \end{picture}
    }
    \caption{%
    (a,b) Temporal variation of coronal hole areas at the north and south poles derived from the SPoCA catalogue. 
    Grey dots denote daily values, orange curves represent smoothed Fourier-filtered data, black curves indicate the long-term background trends, and blue curves show the fitted sinusoidal variations. 
    (c,d) Cross-correlation analysis between the microwave PB peak temperature, absolute magnetic field strength, and coronal hole areas at the north and south poles, respectively.
    }
    \label{fig:3}
\end{figure*}

The SPoCA catalogue utilised to obtain coronal hole (CH) boundaries is provided as polygonal coordinates derived from AIA 193 \AA{} observations. From these coordinates, binary maps were generated in which CH pixels were assigned a value of 1 and all other pixels were assigned 0. For each day, we extracted only the portions of these maps located above $\abs{60^{\circ}}$ latitude to isolate the polar coronal hole regions. 
The total number of pixels with a value of 1 within the extracted regions was computed and used as a proxy for the polar CH area, representing the projected area obtained from the two-dimensional image projection of the spherical solar surface. These daily proxy values are plotted as grey points in \autoref{fig:3}. As in the brightness-temperature analysis, a Fourier filter was applied to remove periodicities associated with the annual and semi-annual variation of the $B_{0}$ angle. A running-average filter with a window of approximately one year was then applied to obtain a smoothed trend, shown by the orange curve in \autoref{fig:3}. We additionally extracted the long-term background variation using the Multiple  Seasonal-Trend Decomposition (MSTL; see \citealt{Bandara2025}) technique, which accommodates both yearly and monthly periodicities present in the data; this trend is plotted in black. Finally, a sinusoidal function was fitted to the original unfiltered values to capture the dominant periodic behaviour, and the resulting fit is overplotted in blue.

\subsection{Temporal and Angular variation of the limb brightening}\label{subsec:t_a_v_l_b}
\begin{figure*}[hbt!]
    \centering
    \begin{subfigure}[b]{\textwidth}
        \centering
        \includegraphics[width=0.96\textwidth]{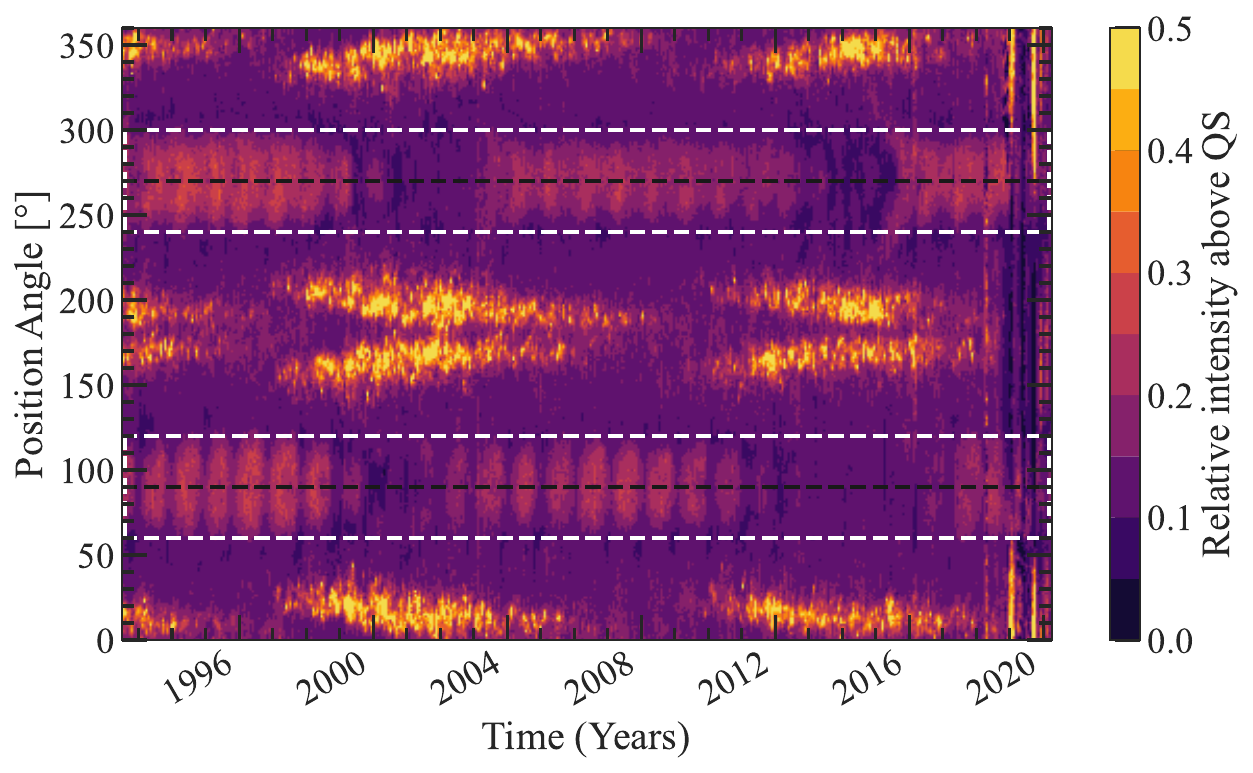}
        \label{fig:4a}
    \end{subfigure}

    \vspace{0.1em}

    \begin{subfigure}[b]{\textwidth}
        \centering
        \includegraphics[width=0.96\textwidth]{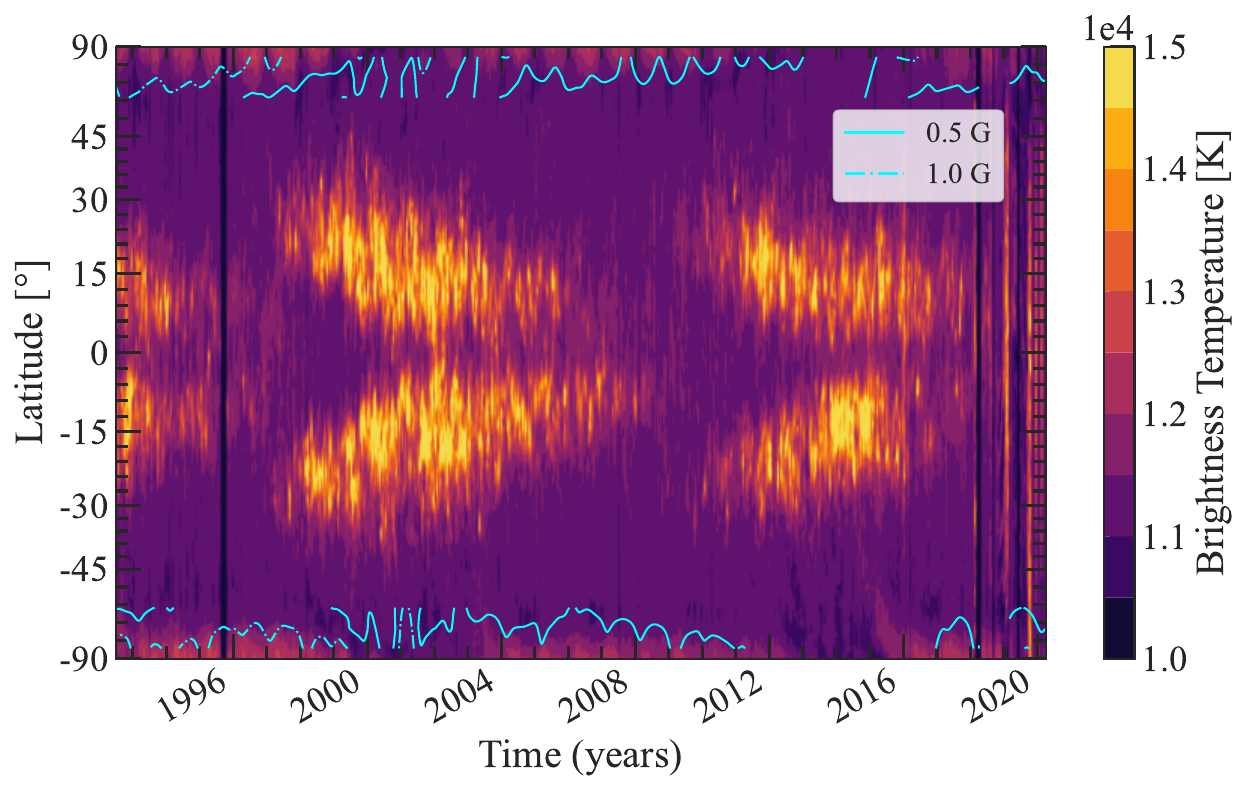}
        \label{fig:4b}
    \end{subfigure}

    \caption{\textbf{Top Panel}: Temporal and Angular variation of the limb brightening. The black dotted horizontal lines show the position of the poles. The white boxes indicate a $\pm30^{\circ}$ window at the poles where the microwave PB is prominent. \textbf{Bottom Panel:} Latitudinal variation of limb brightening as a function of time. Magnetic butterfly contours for particular field strength are overplotted for latitudes $> |55 ^{\circ}|$ in cyan to show the presence of brightenings at similar locations.}
    \label{fig:4}
\end{figure*}

To visualise the angular and temporal variation of the limb brightening, we have opted for the same method used in section \autoref{subsec:t_v_p_b}. However, instead of taking an average of the radial profiles, we stored all the individual peak values for all position angles ($0^\circ$ at the west limb, and proceeding anticlockwise in $1^\circ$ increments), for all available daily datasets of the 17 GHz band of NoRH. By placing all the peak values for all position angles along the y-axis and plotting them for subsequent daily data along the x-axis, we obtain a map as shown in the top panel of \autoref{fig:4}.
By averaging the peak temperature values at equal angular distances from the pole (for example, at $95^{\circ}$ and $85^{\circ}$), we obtain a representative peak temperature for that latitude (e.g., $85^{\circ}$ N in this case). Repeating this procedure for all latitudes and all daily observations yields the plot shown in the bottom panel of \autoref{fig:4}. Magnetic butterfly contours derived from Wilcox Solar Observatory polar data, corresponding to field strengths of 0.5 and 1 G, are overplotted in cyan for latitudes $> |55^{\circ}|$ to highlight the spatial correspondence of brightenings. Prior to contour extraction, the data are smoothed using a Gaussian filter applied exclusively along the temporal direction (kernel width of 100 pixels) to suppress small-scale fluctuations. 

\section{Results and Discussions} \label{sec:results}
\subsection{Long-term behaviour of the microwave PB and its association with PCH and the polar magnetic field}

From the plots in \autoref{fig:2}(a)-(b), we calculated the mean of the excess intensity (temperature) over the quiet sun at the poles to be around $15( \pm 4 )\%$ in case of North pole, and for south pole, the value is around $16( \pm 4 )\%$, which comes around $1000-2000$ K excess temperature (while considering $1\sigma$ above the mean. However, $T_{b}$ of the microwave PB patches can be as high as 3000 K or more; see \autoref{fig:2}). The period of the brightness variation, when fitted with a sine curve (blue dashed curve in \autoref{fig:2}(a)-(b)), is found to be 11.7 years for the north pole and approximately 11.4 years for the south pole. Around July 2000 (4 July 2000 $\pm$ 27 days), the north pole shows its minimum (the mean date and associated uncertainty are estimated using a residual-based stationary bootstrap method, wherein the residual component of the STL decomposition is resampled, the trend is recomputed for each realization, and the distribution of the resulting minimum locations is used to estimate the mean and $1\sigma$ uncertainty; see \citealt{Hardle2003, Bergmeir2016,Petropoulos2018}). In contrast, the south pole minimum occurs later, around July 2001 (8 July 2001 $\pm$ 61 days).

For the subsequent cycle, the north pole does not exhibit a well-defined sharp minimum; instead, it enters an extended minimum phase (plateau). The onset of this plateau is estimated to occur around August 2011 (13 August 2011 $\pm$ 73 days), and it persists until approximately August 2015 (13 August 2015 $\pm$ 47 days), after which the trend begins to rise. In contrast, the south pole shows a well-defined minimum at 28 November 2013 ($\pm$ 26 days), without evidence of a comparable plateau phase.

These results indicate a clear North–South (N–S) asymmetry in the temporal evolution of solar polar brightness, consistent with earlier studies (e.g., \citealt{Gopalswamy2012, Hathaway2016, Janardhan2018}). A similar asymmetry is also evident in the long-term variation of the PCH area shown in \autoref{fig:3}(a)-(b). We further compare the temporal evolution of microwave PB peak temperature and PCH area over the interval March 2010–December 2017 (\autoref{fig:3}(c)-(d)). We find that the PCH area begins to increase from its minimum phase prior to the rise in microwave PB peak temperature for both poles. 
The Spearman rank correlation coefficient ($\rho$) between these two quantities was 0.67 for the North Pole and 0.63 for the South Pole (with $p << 0.001$ in both cases), indicating a strong correlation between them. We also calculated the correlation coefficient between the trends of microwave PB and the polar absolute magnetic field strength for the poles (\autoref{fig:3} c,d). For both poles, $\rho$ is very high ($\sim 0.9$ with $p << 0.001$ for both poles), suggesting a very strong correlation between these two quantities.

\noindent
\subsection{EUV counterparts of the individual microwave PB patches}

\begin{figure*}[ht!]
    \begin{subfigure}{\textwidth}
          \begin{interactive}{animation}{All_images/For_ppr/Rad_contour_over_AIA_newww.mp4}
            \begin{overpic}[width=\linewidth]{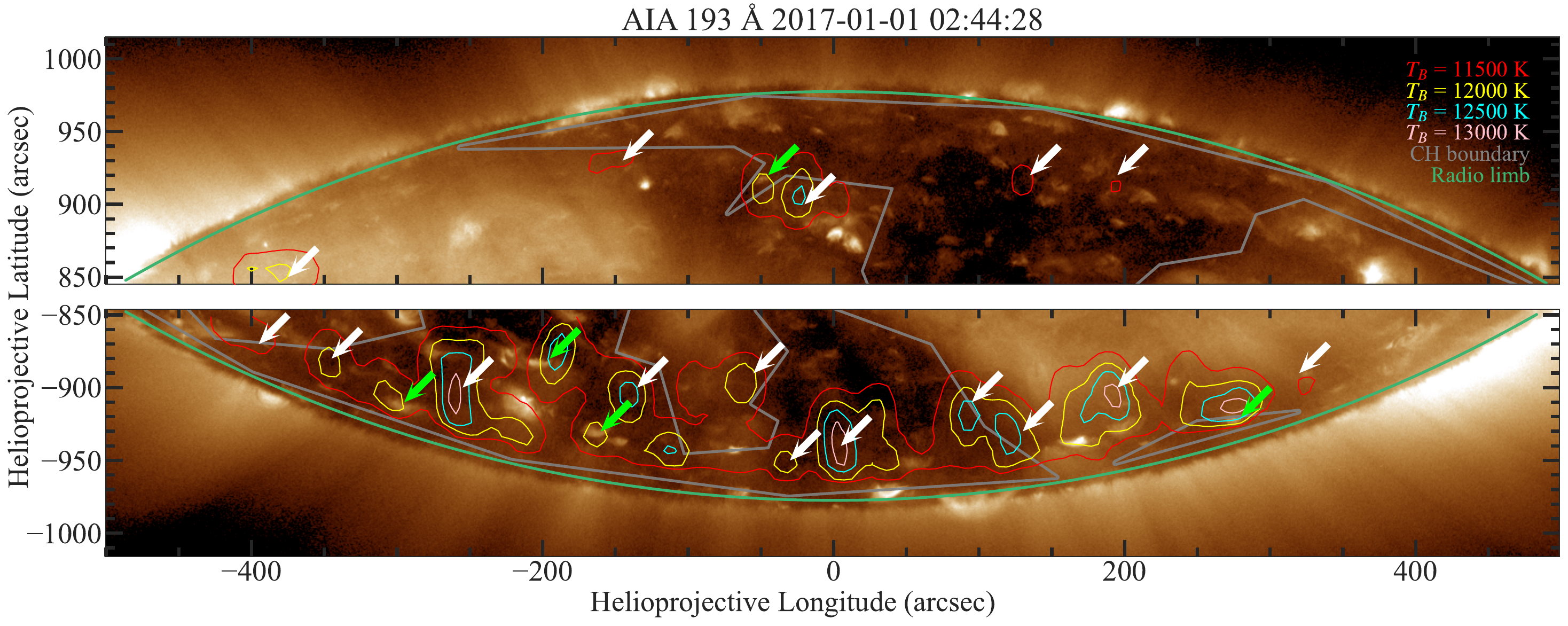}
              \put(97,39){\textbf{(a)}}
            \end{overpic}
          \end{interactive}
        \end{subfigure}
        
    \vspace{0.4cm}
    
    \begin{subfigure}{0.48\textwidth}
        \centering
        \begin{overpic}[width=\linewidth]{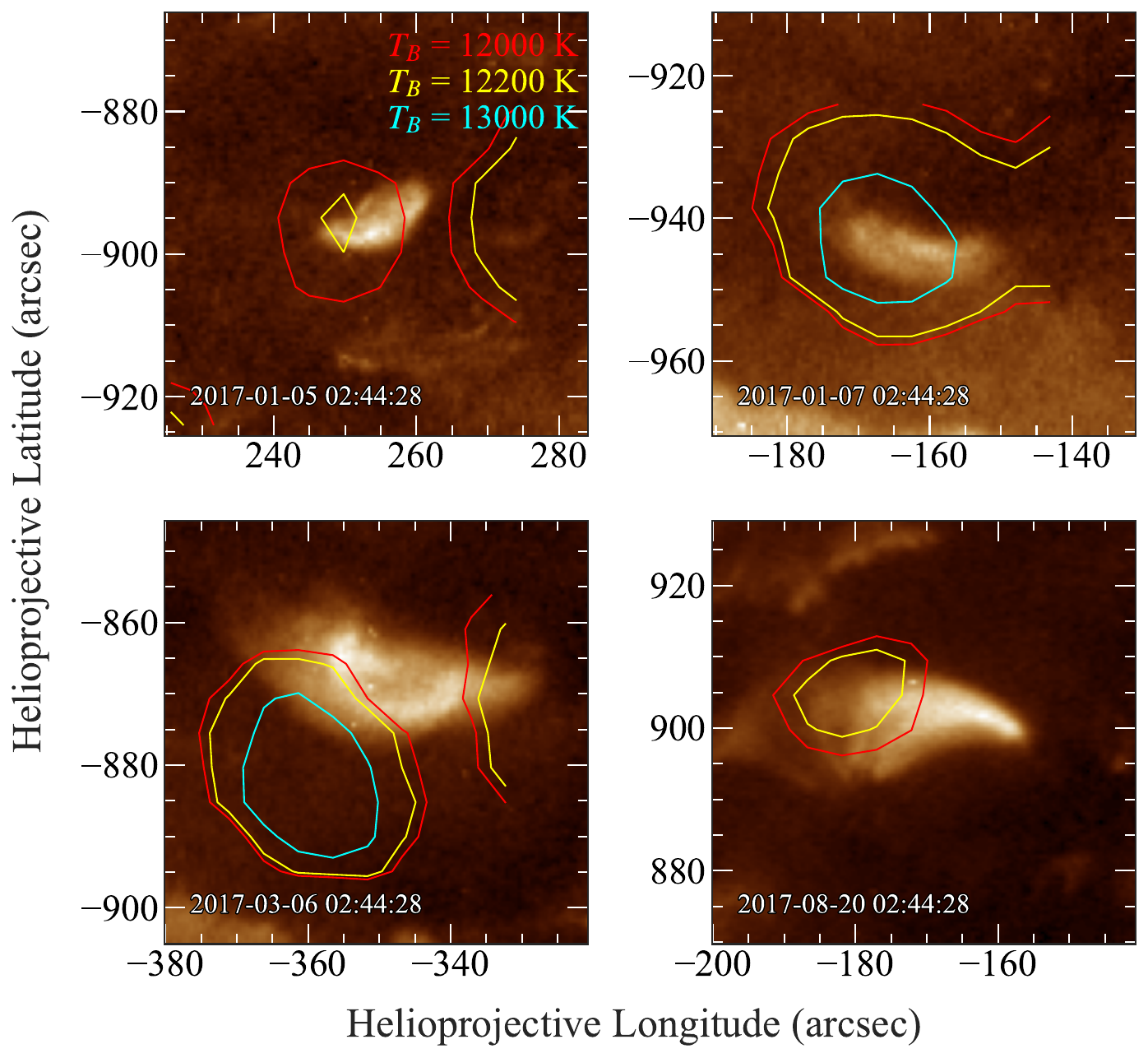}
            \put(95,93){\textbf{(b)}}
        \end{overpic}
        \label{fig:5b}
    \end{subfigure}
    \hfill
    \begin{subfigure}{0.48\textwidth}
        \centering
        \begin{overpic}[width=\linewidth]{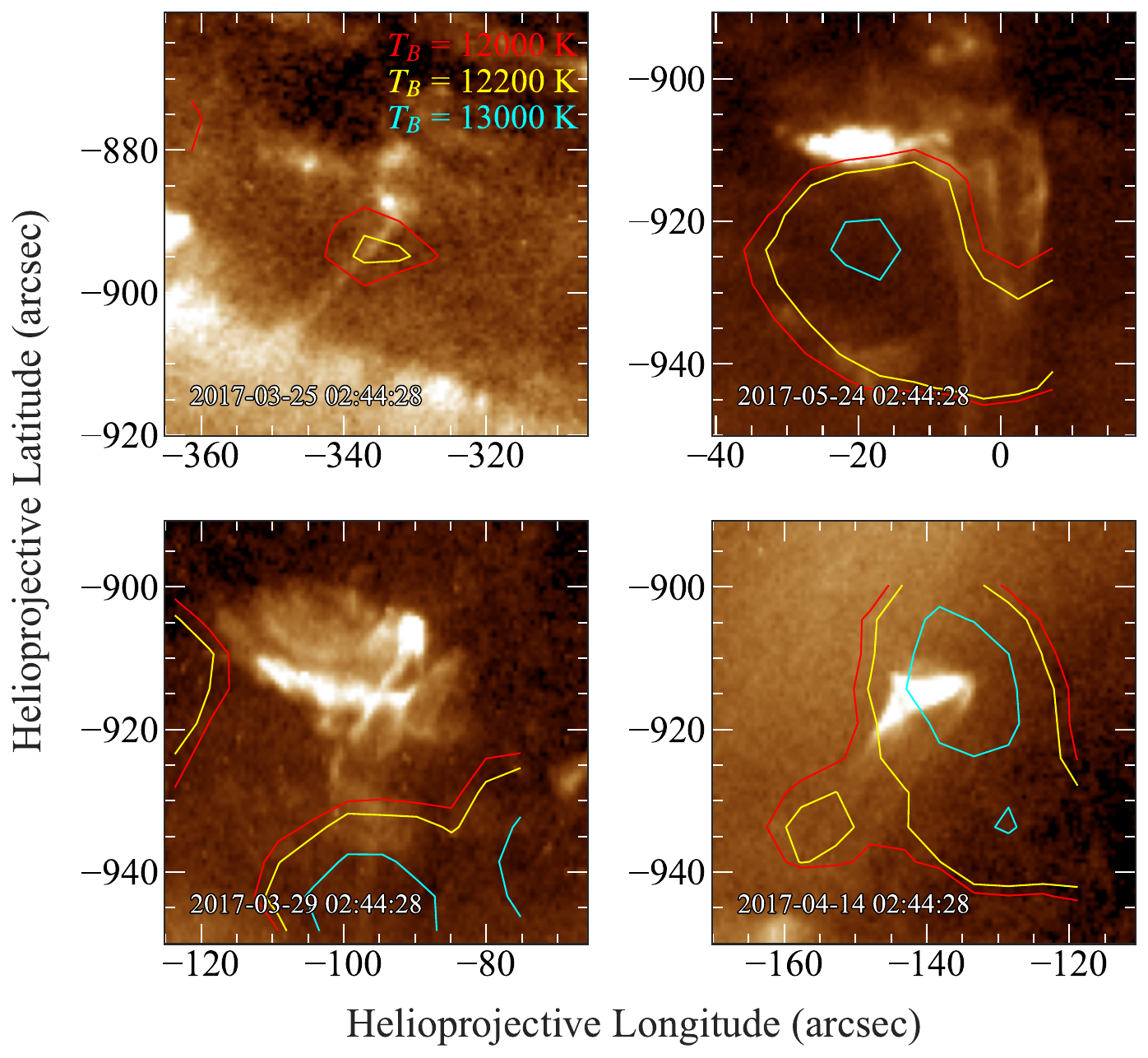}
            \put(95,93){\textbf{(c)}}
        \end{overpic}
        \label{fig:5c}
    \end{subfigure}

    \caption{(a) NoRH 17 GHz microwave PB contours being overplotted on AIA 193 \r{A} data for North (upper panel) and South (lower panel) poles. The green line shows the position of the limb in NoRH data, and the grey line shows the coronal hole boundary from the SPoCA code. The lime arrows indicate microwave PB contours associated with EUV activity, while the white arrows mark regions with no corresponding EUV coronal emission. An animation of this figure is available in the online version of the journal. The animation presents daily EUV images for 2017 with the corresponding microwave polar brightening (MPB) contours overplotted. (b) Some examples of NoRH 17 GHz microwave PB contours overlapping with the AIA 193 \r{A} small-scale loops. (c) Some examples of NoRH 17 GHz microwave PB contours overlapping with the  AIA 193 \r{A} jets. The contour levels are described in the image legends. 
}
    \label{fig:5}
\end{figure*}

To further examine the co-spatial occurrence of microwave PBs and PCHs, we overlaid microwave PB equal temperature contours on AIA 193 \r{A} images of the polar regions using daily observations from 2017 (see the \autoref{fig:5}(a) and the accompanying animation in the online journal). The coronal hole regions identified by the SPoCA algorithm are marked in grey; however, some dark regions that are likely part of the coronal hole are not captured by the algorithm. We find that most microwave PB patches are located within the coronal hole boundaries, although a few exceptions are present (\autoref{fig:5}(a)).

By comparing the NoRH 17 GHz microwave data with the AIA 193 Å observations, we identify three main types of EUV counterparts associated with the microwave PB patches: (i) coronal bright points (small-scale coronal loops), (ii) coronal jets (both standard and blow-out types), and (iii) regions showing no clear EUV brightenings. These cases are discussed in the following subsections.

For January 2017, an extensive manual search for EUV counterparts of the microwave PB patches yielded 85 events exhibiting either category (i) or (ii) association, with only three cases falling into category (ii), and just one event associated with a plume-like structure.
We note that an association was counted only when the EUV brightness was visually significant compared to the surroundings and spatially coincident with, or in close proximity to, the microwave PB contours. It is likely that the number of jets identified here is underestimated, as the analysis is based on single timestamps. A temporal investigation of the EUV data would provide a more accurate estimate.
While thermal bremsstrahlung emission is expected to be co-spatial with EUV-emitting plasma, several factors can introduce apparent offsets, such as projection effects and foreshortening near the solar poles, differences in formation heights and temperature sensitivity between microwave and EUV diagnostics, and uncertainties in image co-alignment and spatial resolution can all lead to small positional mismatches. A more robust assessment will require automated identification methods combined with time-resolved analysis.
In future work, we plan to employ automated methods to identify such co-spatial associations and to investigate their temporal evolution.

\subsubsection{Coronal bright point association}
We identify some cases in which the microwave equal-temperature contours spatially coincide with the compact EUV bright structures observed in the AIA 193 \r{A} images. These structures have typical length scales of 15\arcsec-40\arcsec and majority of them are composed of multiple small-scale loops. Their morphology and characteristic sizes are consistent with those of coronal bright points (CBPs; \citealt{Madjarska2019}). Four representative examples, where the microwave bright contours overlap CBPs, are shown in \autoref{fig:5}(b). Also see the microwave PB contours marked with lime arrows in \autoref{fig:5}(a).

CBPs are thought to form either through `chance' encounters (term coined by \citealt{Harvey1985} to address the encounter between two pre-existing loops) between pre-existing opposite-polarity magnetic loops (termed as cancelling magnetic features or CMFs in the Converging Flux Model by \citealt{Priest1994} ) or via the emergence of magnetic flux from the photosphere \citep{Golub1977}. They are commonly located near majority-polarity open kG magnetic patches, which have the same polarity as the global polar field \citep{Prabhu2020} and are associated with polar faculae \citep{Tsuneta2008,Kaithakkal2013,Narang2019}. Except for the first case shown in \autoref{fig:5}(b), we also detect nearby open-field plume structures in the remaining three events, most clearly visible in the corresponding AIA 171 \r{A} images. \citet{Huang2026} recently proposed that microwave bright points are associated with such open-field plume structures, based on the absence of small-scale EUV brightenings in AIA/SDO 171 \r{A} observations (Nevertheless, we found the existence of a CBP at the base of one plume among the three bright patches they studied). 
In our analysis, in cases where nearby plume structures are present, the subset of events with CBPs associated with microwave bright patches suggests that the microwave emission tends to be located closer to the CBPs, while still occurring in the vicinity of the plume structures. 
Given their compact spatial extent, this may suggest that the microwave emission could, in some cases, be associated with compact magnetic features such as CBPs, in addition to extended plume structures. We note that previous studies (see \citealt{Wang1998,Poletto2015}) have suggested that polar plumes themselves may originate from magnetic reconnection between open field lines and small-scale loops; in our case, these may correspond to open field lines rooted in kG patches and the CBPs, respectively.

\subsubsection{Coronal jet association}
In addition to small-scale loops, we also observed eruptions of coronal jets at the locations of some microwave PBs. Both standard and blow-out jets are identified, with four representative examples shown in \autoref{fig:5}(c). The observed jets may also be linked to underlying CBPs whose fine structure is not clearly resolved in the AIA/SDO data. Magnetic reconnection between newly emerging fields at CBP sites and the open kG field lines can drive such jet activity \citep{Shimojo2009}. We did not find conclusive evidence for the existence of plumes near the region where the jet occurs.

\subsubsection{Bare region}
We also identified some microwave bright patches that do not show any obvious EUV coronal activity (see the microwave PB contours marked with white arrows in \autoref{fig:5}(a)). In \autoref{fig:5}(a), we found 5 regions with association, whereas 15 contours had no clear association. We also checked for January 29, 2017, and there the ratio of association to non-association was 1:2. In a comparative study of upper-photospheric and coronal structures in CBPs, \citet{Shimojo2009} found that coronal activity in polar coronal holes tends to occur only when emerging minority-polarity flux near kG patches attains sufficiently strong magnetic fields in the lower atmosphere. A similar scenario may explain the absence of EUV signatures in the coronal channel for some of the microwave PBs observed in our study.

\subsection{Discussion on the long-term variation of the CBPs}
\citealt{Sattarov2010} classified coronal bright points into two categories based on their peak intensity in EIT/SOHO 195 \r{A} images: bright CBPs, with maximum intensities exceeding 200 DN, which are predominantly associated with active regions, and dim CBPs, with maximum intensities below 200 DN, corresponding mainly to quiet-Sun regions. \citealt{Minenko2021} reported that the total number of coronal bright points (CBPs) at mid-latitudes ($\pm45^\circ$–$55^\circ$) is anti-correlated with the sunspot cycle. In the same study, dim CBPs were found to be strongly anti-correlated with sunspot number, whereas the number of bright CBPs increases toward solar maximum (see Table 1 of the paper). Related results were reported by \citealt{Karachik2025}, who found a positive correlation between total bright CBPs and sunspot number and a negative correlation for total dim CBPs; variations in the total CBP population were the same as those in the dim CBP component. To our knowledge, no dedicated analysis has been carried out for the polar caps; however, these results suggest that changes in the number of compact brightenings such as CBPs over the solar cycle could also influence the variability of microwave polar brightenings, in addition to the background contribution associated with coronal hole area. In this context, it is important to note that the CBPs considered in the present study correspond to the bright CBP population, characterized by small-scale loop structures (see \citealt{Minenko2021, Karachik2025}). 



\section{Summary and Conclusions}
The main findings of this study are summarised as follows.

(1) We find an excess temperature of $15 \pm 4\%$ at the north pole and $16 \pm 4\%$ at the south pole, consistent with earlier results (e.g., \citealt{Selhorst2003}). The long-term evolution of microwave PB shows a clear hemispheric asymmetry. The dominant periodicities differ between the poles (11.7 years for the north and 11.4 years for the south), and the epochs of minimum brightness occur at distinct times (July 2000 for the north pole and July 2001 for the south pole; see \autoref{fig:2} a,b). A similar hemispheric asymmetry is evident in the temporal behaviour of coronal hole (CH) area (\autoref{fig:3} a,b). 

(2) We find a strong correlation between the temporal variation of microwave PB peak temperature and the polar magnetic field strength at both poles, with Spearman rank correlation coefficients of 0.94 in the north and south (\autoref{fig:3} c,d; see also the right panel of \autoref{fig:4}). Additionally, the temporal evolution of microwave PB peak temperature exhibits a strong correlation with the polar CH area, with correlation coefficients of 0.67 in the north and 0.63 in the south (\autoref{fig:3} c,d). Although the CH area may be underestimated due to limitations of the SPoCA automated detection algorithm, the correlation remains strong, indicating a robust relationship between microwave PB peak temperature and CH evolution. This result extends earlier findings that used PFSS-extrapolated polar magnetic fields as a proxy for coronal hole evolution \citep{Fujiki2019}, by directly employing the observed CH area as an additional, independent measure. The consistency between these approaches further reinforces the close coupling between coronal holes and the polar magnetic field. 

(3) A key result of this study is the identification of EUV counterparts to individual microwave bright patches. The physical reality of these microwave bright patches or compact microwave sources (sources similar to those highlighted by the yellow, sky blue, and pink contours in the southern hemisphere in \autoref{fig:5}(a)) has been a subject of extensive investigation, notably addressed by \citet{Nindos1999}, who conducted a detailed analysis of image deconvolution issues using the Nobeyama Radioheliograph (NoRH). By comparing different algorithms, they demonstrated that while the standard CLEAN and Steer CLEAN methods can artificially fragment diffuse emission into ``blotchy'' artifacts, the Maximum Entropy Method (MEM) is better suited to recovering sources that are likely physical in nature, and presented examples of such compact features. Consistent with this, \citet{Huang2026} employed independent reconstruction methods using CLEAN and Steer, and were also able to recover compact bright patches, further supporting their physical reality. While \citet{Nindos1999} reported that many compact features are short-lived (on the order of minutes) and often lack clear EUV counterparts, suggesting either low-temperature origins or sensitivity limitations, later work by \citet{Nitta2014} argued more skeptically. They suggested that most of the compact microwave sources were simply CLEAN artifacts and did not significantly contribute to the long-term variability of microwave polar brightenings (PBs), attributing that variability instead to diffuse emission (emission similar to the red contours in the southern hemisphere in \autoref{fig:5}(a)). However, our analysis and results provide a different interpretation. While we acknowledge the presence of diffuse brightening (and its possible contribution to the long-term variation of the microwave PB), our findings demonstrate a clear physical basis for the compact microwave sources (or microwave bright patches). By overlaying microwave contours on AIA 193~\AA\ images (\autoref{fig:5}), we find that most microwave bright patches are located within coronal hole regions, with a small fraction lying just outside the detected CH boundaries. At EUV wavelengths, many of these patches are associated with coronal bright points (CBPs), appearing as small-scale loop systems. \citet{Huang2026} proposed an association between microwave bright points and nearby open-field plume structures based on the absence of EUV brightenings in AIA 171~\r{A} data. In our analysis, particularly in cases where plume structures are present, the microwave bright patches are often found closer to CBPs, while still occurring in the vicinity of plumes. Given their compact spatial extent, this may suggest that the microwave emission could, in some cases, be associated with compact magnetic features such as CBPs, in addition to extended plume structures. Polar plumes may nevertheless originate from magnetic reconnection between open kG field lines and CBPs for those cases. 
In several cases, both standard and blow-out coronal jets are observed to originate adjacent to or co-spatial with the microwave bright patches, likely representing byproducts of reconnection between open field lines and a bipolar region (which may or may not be a CBP) at spatial scales below the resolution of AIA. The absence of coronal EUV counterparts for some microwave bright points in \autoref{fig:5}(a) is likely attributable to the weak magnetic fields of emerging bipoles associated with CBPs.

As a likely origin of microwave PBs, enhanced free–free emission is expected to dominate, while gyrosynchrotron emission may also contribute in eruptive, compact sources such as CBPs \citep{Alissandrakis2025}. Gyroresonance emission, however, is unlikely to play a significant role, as the magnetic field strengths in polar coronal holes are generally too weak \citep{Dulk1985, Kim2017}. Our analysis suggests that variations in the population of compact brightenings, particularly bright CBPs, may contribute to the long-term variability of microwave polar brightenings, in addition to the background component (or diffuse emission) associated with coronal hole area. A more definitive understanding of the physical mechanisms responsible for these compact microwave sources will require dedicated studies of their temporal evolution using coordinated EUV and microwave observations. Future high-resolution microwave measurements, particularly with instruments such as ALMA, will be valuable for probing the coexistence of EUV and microwave signatures associated with CBPs and jets, thereby providing stronger constraints on the underlying emission mechanisms in polar coronal holes.


\begin{acknowledgments}
RB acknowledges the financial support received from the DST-INSPIRE Fellowship Programme  DST/INSPIRE Fellowship/2021/IF210402). The computational resources utilised in this study were provided by ARIES. AK acknowledges the ANRF Prime Minister Early Career Research Grant (PM ECRG) program. The data utilized in this work were acquired from the Nobeyama Radio Observatory (NRO) database. The authors are grateful to the observers at NRO for providing the data. The ISEE Database for High-Cadence Microwave Images of Solar Flares Observed with Nobeyama Radioheliograph was developed by the Center for Heliospheric Science, Institute for Space-Earth Environmental Research (ISEE), Nagoya University.
\end{acknowledgments}

\begin{contribution}
AK conceptualised the project. RB performed the data analysis and wrote the majority of the manuscript. VP contributed to the discussions and reviewing the manuscript. SR contributed to the initial data analysis and reviewed the manuscript. DP contributed to the initial stages of the data analysis. MVSK helped in solving the reviewer comments and finalizing the paper. All authors contributed to the discussions and summary in the manuscript. All authors reviewed and approved the final manuscript.
\end{contribution}

\facilities{}
Nobeyama Radio Heliograph/ NSO, AIA/SDO, HMI/SDO, MDI/SOHO, Kitt Peak Observatory

\software{}
Sunpy, Matplotlib, Scipy, Statmodels

\appendix

\section{Seasonal Trend Decomposition by LOESS}\label{sec:appendixA}
\noindent
STL (Seasonal–Trend decomposition using LOESS) employs LOESS, a non-parametric regression method that fits smooth curves locally, rather than assuming a global linear or polynomial model. For each point $x_0$ in a dataset, LOESS defines a neighbourhood through a chosen span (fraction or number of nearby points), and assigns weights using the tricube kernel
$d = (x - x_0)/h$ and $w(x) = (1 - |d|^3)^3$, and performs a weighted least-squares fit
$\sum_i w_i [y_i - f(x_i)]^2$.
Repeating this for all points produces a smooth, locally adaptive fitted curve.

\begin{itemize}

\item In STL, LOESS is repeatedly applied to separate a time series into \textit{seasonal}, \textit{trend}, and \textit{residual} components.  
A known seasonal period $P$ is required (e.g., 12 for monthly yearly data, 36 for patterns repeating every three years, or 365 for daily annual data).

\item \textbf{Seasonal component:}  
The data are split into subseries containing observations at the same position within each seasonal cycle (e.g., all January 1 values, all January 2 values, and so on).  
Each subseries is smoothed using LOESS, and the resulting estimates are smoothed again across adjacent seasons, allowing the seasonal pattern to change smoothly over time.

\item \textbf{Deseasonalization:}  
\text{Deseasoned data} = \text{Original data} - \text{Seasonal component}.

\item \textbf{Trend component:}  
A LOESS fit with a much larger span is applied to the deseasoned data to capture long-term, low-frequency variations.

\item \textbf{Residual:}  
\text{Residual} = \text{Original data} - \text{Seasonal component} - \text{Trend component}.

\end{itemize}

\noindent
These steps are iterated, alternately refining the seasonal and trend estimates until convergence, allowing STL to handle data where the seasonal pattern slowly evolves. In this work, we have utilised the \texttt{STL} module from the \texttt{statsmodels} Python library to perform STL analysis on our dataset.

\bibliographystyle{aasjournalv7}




\end{document}